\documentclass[10pt,prc,aps]{revtex4}
\draft

\usepackage{graphicx}
\usepackage{subfigure}
\usepackage{color}
\usepackage{mathtools}
\begin{document}

%\title{The far-reaching features of low-density neutron matter} 
\title{Neutron matter: The unitary limit and its far-reaching impact}  
\author{            
Francesca Sammarruca\footnote{fsammarr@uidaho.edu} }                                                         
\affiliation{ Physics Department, University of Idaho, Moscow, ID 83844-0903, U.S.A. 
}

\date{\today} 

\begin{abstract}
We explore low-density neutron matter and its behavior in proximity to the unitary limit. To that end, we construct unitary nucleon-nucleon potentials with infinite $^1$S$_0$ neutron-neutron scattering lengths. We discuss the Berstch parameter  in relation to results from ultra-cold atomic gases. Unitarity as a constraint for neutron matter and the symmetry energy has been discussed in the literature. We revisit some of those arguments and emphasize the relevance of keeping a firm link with low-energy nuclear physics for robust predictions of neutron-rich systems. Our predictions are obtained from realistic few-nucleon forces based on chiral effective field theory at N$^3$LO.
\end{abstract}
\maketitle

\section{Introduction} 
\label{Intro} 

The physics of neutron matter (NM) spans a broad range of densities. At low density, it approaches universal
behavior as a consequence of the large neutron-neutron scattering length in the spin-singlet channel. Around normal nuclear density, 
 it is an appropriate laboratory to study neutron-rich nuclei, and, at even higher densities, it strongly constrains the physics of neutron stars. Although an idealized system, neutron matter also provides unique opportunities
to test nuclear forces, because all low-energy constants appearing in the three-neutron forces are
predicted at the two-neutron level.

Recently, we presented our latest results for the equation of state (EoS) of NM and the properties of the symmetry energy, based on
 high-quality nuclear forces constructed within the framework of chiral effective field theory (EFT)~\cite{SM21,FS22}.
Since the past several years, chiral EFT is generally recognized as the most fundamental approach  for developing nuclear interactions: it provides a systematic way to construct nuclear two- and 
many-body forces on an equal footing~\cite{Wei92} and allows
to assess theoretical uncertainties through an expansion controlled by an organizational scheme known as
``power counting"~\cite{Wei92}. Furthermore, chiral EFT maintains consistency with the underlying fundamental theory of strong interactions, quantum chromodynamics (QCD), through the symmetries and
symmetry breaking mechanisms of low-energy QCD.

This work is a specific investigation of dilute NM within the same framework. Predictions are based on our most recent EoS~\cite{SM21, FS22} derived from two- and three-nucleon chiral forces consistently at N$^3$LO.

A  strong motivation for studying the physics of low-density NM is its impact on the inner crust of
neutron stars, where the nucleon density varies approximately from 10$^{-3}$ to 0.08 fm$^{-3}$. In that region, matter consists of a mixture
of very neutron-rich nuclei (arranged in a Coulomb
lattice), electrons, and a superfluid neutron gas.

At low densities, the neutron-neutron interaction is dominated by the attractive component of the $^1$S$_0$ partial wave, which, 
although insufficient to bind two neutrons, leads to a virtual dineutron state and thus the large
neutron-neutron scattering length in this channel, $ a_s$ = -18.9(4) fm~\cite{Chen+}. Therefore, even if the average distance between two neutrons  is much larger than the effective range in the $^1$S$_0$ neutron-neutron
interaction, $r_e$ = 2.75(11) fm~\cite{Miller+}, NM is still a strongly correlated system. Note that $k_F^{-1}$, being proportional to the cubic root of the unit volume, can be taken as an approximate measure of the average distance between fermions.

A unitary Fermi gas is an idealized system of fermions with a
zero-range interaction having an infinite (negative) scattering
length. All properties of this interacting gas are simply proportional
to the corresponding ones in the non-interacting system at the same density.

The model of a dilute quantum gas can describe diverse systems and can be applied in different areas of physics. At the low temperatures required to observe quantum phenomena in a dilute gas, the type of fermions and the exact form of the interaction become unimportant for the macroscopic properties of the system. 
For these reasons, results from the field of cold gases can give insight into systems such as NM at subnuclear densities. This unique feature of interacting Fermi gases -- unitarity -- can be generalized to any system of fermions subjected to mutual interactions with diverging scattering lengths. The unitary limit was first introduced in 1999 by Bertsch~\cite{Bert}, who proposed to model low-density neutron matter as a Fermi gas where
\begin{equation}
 r_e << k_F^{-1} << |a_s| \; ,  \; \; \; \; \; \mbox{implying:} \; \; \; \; \; k_F |a_s| >> 1 \; .
\end{equation}
 Universality of unitary systems implies that the ground state energy at the unitary limit should be given by
\begin{equation}
 E(k_F) = \xi E_{FG}(k_F) \; ,
\label{xi}
\end{equation}
 where $\xi$ is known as the Bertsch parameter and $E_{FG}(k_F)$ is the energy of the corresponding non-interacting Fermi gas. Measurements of the Berstch parameter with ultra-cold atomic gases reported values ranging from $\sim$0.36 to $\sim$0.51~\cite{cold1, cold2, cold3,cold4,cold5, cold6}.

 In the physics of cold atoms near the Feshbach resonance~\cite{Feshb}, external magnetic fields allow the scattering length to be tuned arbitrarily. As mentioned above, in the regime of divergent scattering lengths,  the gas is both dilute (the range of the interatomic potential is much smaller than the interparticle distance) and, at the same time, strongly
interacting (the scattering length is much larger than the interparticle distance).
 Thus, the only scale in the unitary gas (at zero tempearture) is the Fermi momentum of the system -- hence, the universal rescaling of bulk observables.

Unitary Fermi gases have been obtained in the laboratory with ultra-cold trapped alkali atoms, where the effective range of the interaction is approximately equal to $10^{-4} k_F^{-1}$.  In these experiments, it is possible to tune the inter-atom interaction from a weak to a strong coupling regime, and thus to explore the crossover from Bardeen--Cooper--Schrieffer (BCS) pairing with weakly attractive ($a_s < 0$) Cooper pairs to the Bose--Einstein condensation (BEC) of bound dimers ($a_s > 0$)~\cite{40,41,42}. 
Theoretical predictions for $\xi$ range from 0.3 to 0.7~\cite{Baker, 19,20,21,22,23,24,28}. Variational Monte Carlo~\cite{29}, Green's function Monte Carlo~\cite{30}, and Brueckner--Hartree--Fock (BHF)~\cite{31} calculations of the EoS of low-density NM tend to cluster around $\xi$= 0.5.

After a brief review of the theoretical framework (Section~\ref{TF}), we discuss low-density neutron matter and its behavior at the unitary limit, Section~\ref{uni}. In Section~\ref{uni_a}, we pay special attention to the unitary limit as a constraint for the symmetry energy at normal density. Closely related to the nature of low-density NM as a highly interacting system of fermions is S-wave superfluidity.
In Section~\ref{gap}, we present baseline results for the singlet gap in NM obtained {\it via} the BCS equation.
Conclusions, work in progress, and future prospects are found in Section~\ref{Concl}.

\section{Theoretical framework} 
\label{TF}

The EoS for neutron matter is obtained at the leading-order in the hole-line  expansion---namely, {via} a non-perturbative summation of the particle--particle ladders. 
The single-neutron potentials are computed self-consistently with the $G$-matrix, employing a continuous spectrum.

The interactions we use are derived from 
chiral effective field theory (EFT)~\cite{Wei92}, which provides a path to a consistent development of nuclear forces. Symmetries relevant to low-energy QCD are incorporated in the theory, in particular chiral symmetry. Thus, although the degrees of freedom are pions and nucleons instead of quarks and gluons, there exists a solid connection with the fundamental theory of strong interactions through its symmetries and the mechanism of their breaking.
Chiral EFT employs a power counting scheme in which the progression of two- and many-nucleon forces is constructed following a well-defined hierarchy. This allows for the inclusion of all three-nucleon forces (3NFs) which appear at a given order, thus eliminating the inconsistencies which are unavoidable when adopting meson-theoretic or phenomenological forces. Finally, it provides a clear method for controlling the truncation error on an order-by-order basis. 
Next, we give a  brief summary of the input two-nucleon forces (2NFs) and 3NFs. Additional information, including the LECs we use, can be found in Refs.~\cite{SM21,SNM21}.

The 2NF we apply are from Ref.~\cite{EMN17}, a family of high-quality potentials from leading order (LO) to fifth order (N$^4$LO) of chiral EFT, using a non-local regulator~\cite{SM21}. 
The long-range part of these potentials is tightly constrained by the $\pi N$ low-energy constants (LECs) from the Roy--Steiner analysis of Ref.~\cite{Hofe+, Hofe2}. This analysis is sufficiently accurate to render errors in the $\pi N$ LECs essentially negligible for the purpose of quantifying the uncertainty. With the choice
$\Lambda$ = 450 MeV---which we maintain throughout this paper---the potentials are soft according to the
Weinberg eigenvalue analysis of Ref.~\cite{Hop17} and the perturbative calculations of infinite matter from Ref.~\cite{DHS19}.

Three-nucleon forces first appear at the third order of the chiral expansion (N$^2$LO) of the $\Delta$-less theory, which is the one we apply in this work. At this order, the 3NF consists of three contributions~\cite{Epe02}: the long-range two-pion-exchange (2PE) graph, the medium-range one-pion-exchange (1PE) diagram, and a short-range contact term.

The 3NF at N$^3$LO has been derived in Refs.~\cite{Ber08,Ber11}. The long-range part of 
 the subleading chiral 3NF consists of three topologies: the 2PE topology, which is the longest-range component of the subleading 3NF, the two-pion-one-pion exchange (2P1PE) topology, and the ring topology, generated by a circulating pion which is absorbed and reemitted from each of the three nucleons. 

In infinite matter, these 3NF can be expressed as density-dependent effective two-nucleon interactions as derived in Refs.~\cite{holt09,holt10}. They are represented in  terms of the well-known non-relativistic two-body nuclear force operators and, therefore, can be conveniently incorporated in the usual $NN$ partial wave formalism and the particle-particle ladder approximation for computing the EoS. The effective density-dependent two-nucleon interactions at N$^2$LO consist of six one-loop topologies. Three of them are generated from the 2PE graph of the chiral 3NF and depend on the LECs $c_{1,3,4}$, which are already present in the 2PE part of the $NN$ interaction. Two one-loop diagrams are generated from the 1PE diagram, and depend on the low-energy constant $c_D$. Finally, there is the one-loop diagram that involves the 3NF contact diagram, with LEC $c_E$. The last two sets do not contribute in neutron matter.
 
The in-medium $NN$ potentials corresponding to the long-range subleading 3NFs are given in Ref.~\cite{Kais19} for SNM and in Ref.~\cite{Kais20} for NM. The short-range subleading 3NF consists of: the one-pion-exchange-contact topology (1P-contact), which gives no net contribution, the two-pion-exchange-contact topology (2P-contact), and relativistic corrections, which depend on the $C_S$ and the $C_T$ LECs of the 2NF and are proportional to $1/M$, where $M$ is the nucleon mass. 
Expressions for the in-medium $NN$ potentials corresponding to the short-range subleading 3NFs can be found in Ref.~\cite{Treur} for NM.

\section{Low-density neutron matter and the unitary limit} 
\label{uni}

\begin{figure*}[!t] 
\centering
\hspace*{-1cm}
\includegraphics[width=8.7cm]{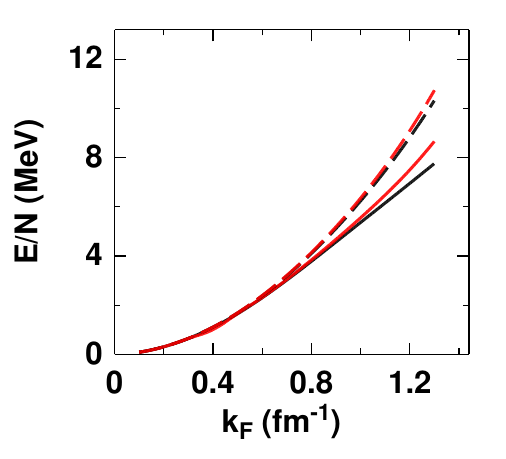}\hspace{0.01in}
\vspace*{-0.5cm}
 \caption{Energy per neutron in NM {\it vs.} the Fermi momentum. For either color (black pair or red pair), the solid curve is obtained with all partial waves (up to $J$=15), while the dashed curve includes only the spin-singlet $S$-wave. The black curves show the result of including only the 2NF, while the red curves are obtained with the inclusion of the full 3NF at N$^3$LO. 
}
\label{fig1ab}
\end{figure*}

Here, we discuss some of the remarkable features of low-density NM.
We begin with identifying the region of $^1$S$_0$ dominance, see  Fig.~\ref{fig1ab}. The largest (neutron) Fermi momentum included in the figure, 1.3 fm$^{-1}$, corresponds to a density of 0.074 fm$^{-3}$, close to one-half of normal nuclear density (taken equal to 0.16 fm$^{-3}$). One may conclude that the $^1$S$_0$ channel is dominant up to about $k_F$ = 1.0 fm$^{-1}$. Furthermore, comparing either black curve with the red of corresponding pattern, we see that the impact of the 3NF at these densities is negligible or minor.

Next, we wish to explore to which extent  NM approaches a unitary gas and how its behavior changes if the scattering length tends to (negative) infinity. For this analysis, 
 we restrict the Fermi momentum to values up to 1.0 fm$^{-1}$ and focus on the $^1$S$_0$ channel. 

\begin{figure*}[!t] 
\centering
\hspace*{1.5cm}
\includegraphics[width=14.5cm]{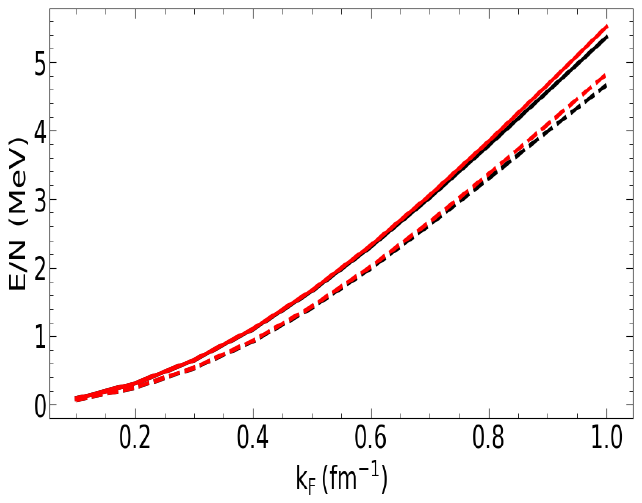}\hspace{0.01in}
\vspace*{-11.0cm}
 \caption{Black curves: energy per neutron {\it vs.} the Fermi momentum using the N$^3$LO(450) potential from Ref.~\cite{EMN17} (solid) or its nearly-unitary counterpart (dash). Red curves: same comparison with CD-Bonn (solid) and its nearly-unitary version (dash).
}
\label{fig2}
\end{figure*}

Starting from the chiral potential at N$^3$LO from Ref.~\cite{EMN17} with cutoff equal to 450 MeV, we constructed an {\it ad hoc} potential that is unitary in $^1$S$_0$, having a scattering length of nearly -10$^4$ fm -- essentially infinity. This comparison is shown 
in Fig.~\ref{fig2}, where the pair of black curves, solid and dashed, are the predictions obtained with the N$^3$LO potential and its unitary version, respectively. 
We also consider one of the popular high-precision meson-exchange potentials of the 90's, the CD-Bonn potential~\cite{CDB}, and constructed a nearly-unitary version for CD-Bonn as well. These are also shown in Fig.~\ref{fig2} as the pair of red curves. In all cases, only the 2NF is included.
Although N$^3$LO and CD-Bonn are based on very different philosophies, accurate reproduction of low-energy free-space data is a stringent constraint for neutron matter~\cite{FS22}, hence the strong similarity between the two solid (or the two dashed) curves. We also note that the
 solid and the dashed curves  (of either color) differ by less than 1 MeV, consistent with neutron matter approaching the behavior of a unitary system at these densities. We emphasize that a change of only a few percent of the leading order contact parameter in $^1$S$_0$ is sufficient to cause very large variations in $a_s$, a sensitivity similar to the one seen when tuning of the interaction with magnetic fields in experiments with cold atoms.

\begin{figure*}[!t] 
\centering
\hspace*{1.5cm}
\includegraphics[width=14.5cm]{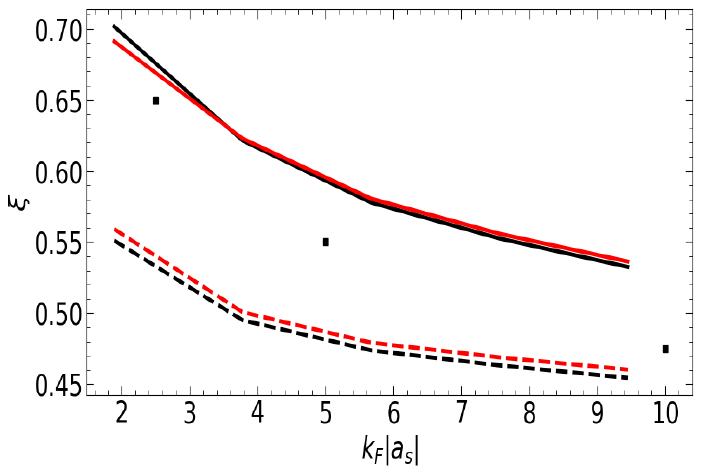}\hspace{0.01in}
\vspace*{-11.5cm}
 \caption{Energy per neutron in units of the free Fermi energy, $\xi$. The solid and dashed black curves are obtained with the original N$^3$LO potential and its unitary counterpart, respectively; The same comparison with the CD-Bonn potential and its unitary counterpart is shown by the red solid and dashed curves. The horizontal scale displays the dimensionless parameter $|a_s k_F|$. The filled squares are results from cold atoms~\cite{Gez08}.
}
\label{fig3}
\end{figure*}

In a discussion of unitary gases, it is insightful to display the ratio of the energy of the interacting gas to the energy of a free Fermi gas at the same density -- the aforementioned Bertsch parameter. 
In Fig.~\ref{fig3}, we show such ratio calculated with the original N$^3$LO potential (solid black) and  its unitary version (dashed black). An analogous description applies to the pair of red curves, but for CD-Bonn.
We observe that the two potentials are impacted in almost exactly the same way by enforcing unitarity in $^1$S$_0$. The nearly-unitary interactions show a clear tendency to join the cold atoms results~\cite{Gez08}, solid squares,  towards the unitary limit (where the Bertsch parameter is defined).

\subsection{Unitarity as a constraint for neutron matter and the symmetry energy}
\label{uni_a}

A few years ago, Tews {\it et al.}~\cite{Tews+} proposed the existence of a lower limit on the energy of NM based on unitarity. With an eye on recent results from electroweak scattering, we wish to revisit that discussion 
for the purpose of emphasising the importance of low-energy constraints for NM and the symmetry energy. 

First, we note that our predictions in Fig.~\ref{fig2} and Fig.~\ref{fig3} are consistent with
\begin{equation}
E_{NM}(\rho) \ge E_{UG}(\rho) \; ,
\label{lb}
\end{equation}
where $UG$ stands for unitary gas. If this is a general bound -- as it seems likely, since higher total energy (that is, less attraction) can be expected in NM as compared to the system with infinitely large neutron-neutron scattering length -- then one can write
\begin{equation}
E_{NM}(\rho) \ge E_{UG}(\rho) = \xi_0 E_{FG}(\rho)  \; ,
\label{lb1}
\end{equation}
where $\xi_0$ is the Bertsch parameter, and $E_{FG}$ is the energy of the non-interacting Fermi gas,
\begin{equation}
E_{FG} = \frac{3 \hbar ^2  k_F^2}{10 m} \; \; \; \; \; \; \; \; \;   k_F = (3 \pi ^2 \rho)^{1/3} \; .
\end{equation}
 Thus, in the parabolic approximation to the symmetry energy, we can write:
\begin{equation}
E_{sym}(\rho) = E_{NM}(\rho) - E_{SNM}(\rho)  \ge E_{UG}(\rho) - E_{SNM}(\rho)   \; ,
\label{lb2}
\end{equation}
where $SNM$ signifies symmetric nuclear matter. 

There exist ``established" expansions of the energy in SNM in terms of the saturation parameters $E_0 = E(\rho_0)$ and the incompressibility $K_0$. The next term in the expansion is the skewness, $Q_0$, which is poorly known. To streamline the notation, we express density in units of saturation density and define $u = \frac{\rho}{\rho_0}$. The expansion of the SNM energy is then written as
\begin{equation}
E_{SNM}(u) = E_{SNM}(u=1) + \frac{K_0}{18}(u-1)^2 + \frac{Q_0}{162}(u-1)^3 + ... \; ,
\label{snm}
\end{equation} 
where $u=1$ at $\rho=\rho_0=0.16 fm ^{-3}$, $E_{SNM}(u=1) = E_0 = -16$ MeV, and $K_0 = 230$ MeV.  In Ref.~\cite{Tews+}, some estimates are made for $Q_0$. Clearly, reliable constraints on these parameters must be available from independent sources. We emphasize, though, that the specific values are not very relevant for the present discussion, which is a qualitative demonstration of how the unitarity constraint propagates. 

From Eq.~(\ref{lb2}) and Eq.~(\ref{snm}), ignoring higher-order terms in the SNM expansion, we have
\begin{equation}
E_{sym}(u) \ge E_{UG}(u) - \big ( E_0 + \frac{K_0}{18}(u-1)^2 + \frac{Q_0}{162}(u-1)^3 \big ) \; .
\label{lb3} 
\end{equation}

\begin{figure*}[!t] 
\centering
\hspace*{2cm}
\includegraphics[width=15.0cm]{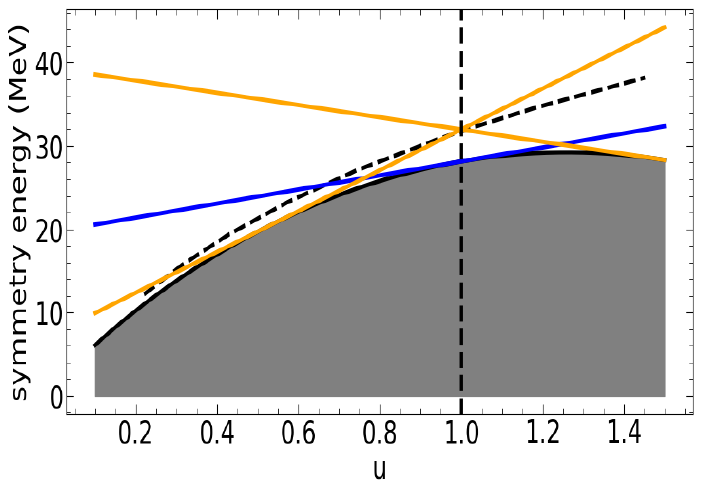}\hspace{0.01in}
\vspace*{-11.9cm}
 \caption{Lower bound for the symmetry energy from Eq.~(\ref{lb3}). The grey area is the excluded region for the symmetry energy. The black dashed vertical line marks the saturation density. The blue line is tangent to the lower-bound curve at $u$ = 1. As an example, a symmetry energy equal to 32 MeV is chosen. Each of the two yellow lines goes through the chosen value and is tangent to the lower-bound curve at a point with $u < 1$ or $u > 1$. The black short-dashed curve shows our predictions.
}
\label{fig4}
\end{figure*}

Equation~(\ref{lb3}) sets a lower bound for the symmetry energy, as shown in Fig.~\ref{fig4}, where the shaded area is excluded. Furthermore, replacing the symmetry energy with its well-know expansion about saturation density,
\begin{equation}
E_{sym}(u) = E_{sym,0} + \frac{L}{3}(u-1)  + \frac{K_{sym}}{18} (u-1)^2 + ... \; \; \; \; \; \; \; \; \; \;  E_{sym,0} = E_{sym}(u=1) \; ,
\label{lb4} 
\end{equation}
one can turn Eq.~(\ref{lb3}) into a constraint for $L$, the slope of the symmetry energy at saturation. For the purpose of this semi-analytical exercise~\cite{Tews+}, we keep one term beyond the leading order in both the expansion of the symmetry energy and the SNM energy:
\begin{equation}
E_{sym}(u) \approx E_{sym,0} + \frac{L}{3}(u-1) \;  ,
\label{lb5} 
\end{equation}
and rewrite Eq.~(\ref{lb3}) as 
\begin{equation}
 E_{sym,0} + \frac{L}{3}(u-1) \ge E_{UG}(u) - \big ( E_0 - \frac{K_0}{18}(u-1)^2 \big ) \; .
\label{lb6} 
\end{equation}
So, for a chosen value of $E_{sym,0}$, we can write, for $u > 1$:
\begin{equation}
  L \ge   \frac{3}{(u-1)} \big (E_{UG}(u) - E_0 - \frac{K_0}{18}(u-1)^2   -E_{sym,0} \big ) \; ,
\label{lb7} 
\end{equation}
and, for $u < 1$:
\begin{equation}
  L \le   \frac{3}{(u-1)} \big (E_{UG}(u) - E_0 + \frac{K_0}{18}(u-1)^2   -E_{sym,0} \big ) \; ,
\label{lb8} 
\end{equation}
thus setting a lower and an upper limit to $L$. This is shown in Fig.~\ref{fig4} by the two yellow lines passing through the chosen value of $E_{sym,0}$ and tangent to the lower-bound contour at two points, one with $u < 1$ and the other with $u > 1$. For this simple example, we find 
\begin{equation}
 -18.7 \mbox{MeV} \le L \le 70.5 \mbox{MeV} \; ,
\label{lb8b} 
\end{equation}
where the negative lower limit is a consequence of the approximations we have applied. Proceeding along these lines, one can then construct a contour in the ($L - E_{sym,0}$) plane. Obviously, unitarity is one of many constraints, from both microscopic theory and experiments, to be considered carefully and vetted for consistency.  The focal point of this discussion is to emphasize the inherent connection between NM and low-energy few-nucleon systems, due to the proximity of few-nucleon systems to the unitary limit.
 An interesting discussion can be found in Ref.~\cite{Kiev+}, where correlations were identified between NM and few-nucleon observables. 

\section{The superfluid singlet gap: baseline results}
\label{gap}

In this section, we address the superfluid gap in the singlet channel. 
A pairing gap is a two-body correlation around the Fermi surface which can manifest itself in any quantum system of fermions. Similar to the formation of Cooper pairs~\cite{bcs} in the presence 
of the attractive phonon-mediated interaction between electrons, pairing in nuclei leads to the formation of a superfluid state that opens a gap at the Fermi surface and reduces the available phase space. Isovector pairing in NM is relevant for the physics of neutron-rich nuclei, especially halo nuclei, and neutron star matter.

To calculate the gap at the simplest level, one uses the BCS (from the theory of electron
superconductivity, now referred to as BCS theory~\cite{bcs}) approximation with the bare two-body potential and free-space single-particle (sp) energies. 
The gap equation then reads, for an uncoupled state,
\begin{equation}
\Delta(k) = -\frac{1}{\pi} \int_0^{\infty} \frac{k'^2V(k,k') \Delta(k')d k'} {\sqrt{ (e(k') - e_F)^2 + \Delta^2(k')}} \; ,
\label{gap_eq} 
\end{equation}
where $e(k)$ is the single-particle energy, $e_F$ is the Fermi energy, $V(k,k')$ is the pairing interaction, and $\Delta(k)$ the unknown gap function, to be determined in a selfconsistent way. Note that the Fermi energy approximates the chemical potential, which should be evaluated selfconsistently with the gap.
We follow the solution method from Ref.~\cite{Khodel96}, of which we summarize the main steps. 

Following the Khodel's method~\cite{Khodel96}, the potential $V(k,k')$ is decomposed as
\begin{equation}
V(k,k') = V_F \phi(k)\phi(k') + W(k,k')  \; ,
\label{V-sep}
\end{equation}
where $V_F = V(k_F,k_F) \ne 0$ and $\phi(k) = V(k,k_F)/V_F$. Note that $\phi(k_F) = 1$ and $W(k_F,k) =0$. Employing these definitions and defining a 
dimensionless gap function through $\Delta(k) = \Delta_F \chi(k)$, Eq.~(\ref{gap_eq}) can be written as two equations:
\begin{equation}
\chi(k) = \phi(k)  -\frac{1}{\pi} \int_0^{\infty} \frac{k'^2W(k,k') \chi(k')d k'} {\sqrt{ (e(k') - e_F)^2 + \Delta_F^2 \chi^2(k')}} \; ,
\label{Khod1}
\end{equation}
and 
\begin{equation}
  -\frac{1}{\pi}  V_F \int_0^{\infty} \frac{k'^2\phi(k')\chi(k') d k'} {\sqrt{ (e(k') - e_F)^2 + \Delta_F^2 \chi^2(k')}}  = 1 \; .
\label{Khod2}
\end{equation}
The advantage of this transformation is that the integrand in Eq.~(\ref{Khod1}) vanishes for $k=k_F$ and thus the nearly singular behavior of the original equation is removed.
The solution scheme~\cite{Khodel96} consists of the following steps: replace $\Delta_F^2 \chi^2(k')$ in the denominator of Eq.~(\ref{Khod1}) by some initial (constant) value, say $\Delta_0$, and solve the resulting (now linear) integral equation with standard matrix inversion techniques to obtain
 a first approximation for $\chi(k)$, 
say $\chi^{(1)}(k)$; replace $\chi^{(1)}(k)$ into Eq.(\ref{Khod2}) to obtain a corresponding solution for $\Delta_F$, $\Delta^{(1)}_F$; repeat the iteration scheme until satisfactory convergence is reached.  
 Equation~(\ref{Khod2}) is perhaps more insightful if written as
\begin{equation}
  \frac{1}{V_F}  + \frac{1}{\pi}
 \int_0^{\infty} \phi(k) \frac{k^2(\phi(k) - \chi(k)) d k} {\sqrt{ (e(k) - e_F)^2 + \Delta_F^2 \chi^2(k)}}  - 
 \frac{1}{\pi}
 \int_0^{\infty} \frac{k^2\phi^2(k) d k} {\sqrt{ (e(k) - e_F)^2 + \Delta_F^2 \chi^2(k)}}  = 0 \; .
\label{Khod3}
\end{equation}
Recalling that $\chi(k_F) = \phi(k_F) = 0$ and arguing that a major contribution to the integral comes from momenta near the Fermi surface, the second integral in Eq.~(\ref{Khod2}) is expected to be nearly insensitive to $\Delta_F$, which facilitate the convergence.

\begin{figure*}[!t] 
\centering
\hspace*{2cm}
\includegraphics[width=18.0cm]{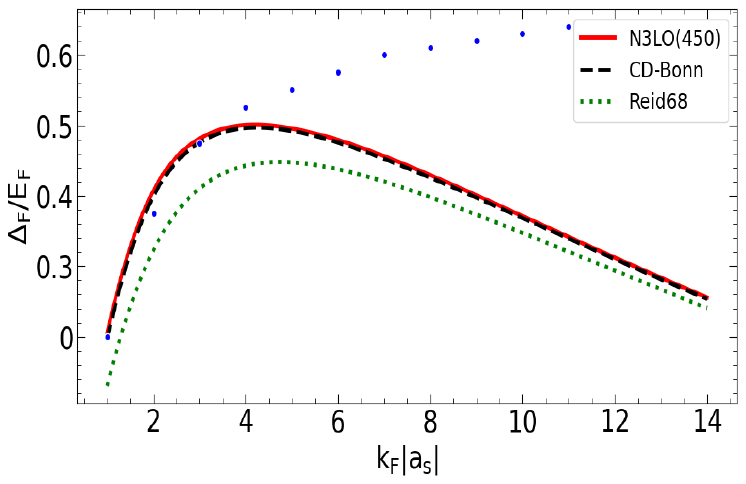}\hspace{0.01in}
\vspace*{-14.5cm}
 \caption{The singlet gap in units of the Fermi energy obtained with: the chiral interaction (solid red), the meson-exchange CD-Bonn potential (black dash), the local phenomenological Reid potential (dotted green). The blue dots are results from QMC calculations of cold atoms. }
\label{fig5}
\end{figure*}   

\begin{figure*}[!t] 
\centering
\hspace*{2cm}
\includegraphics[width=15.0cm]{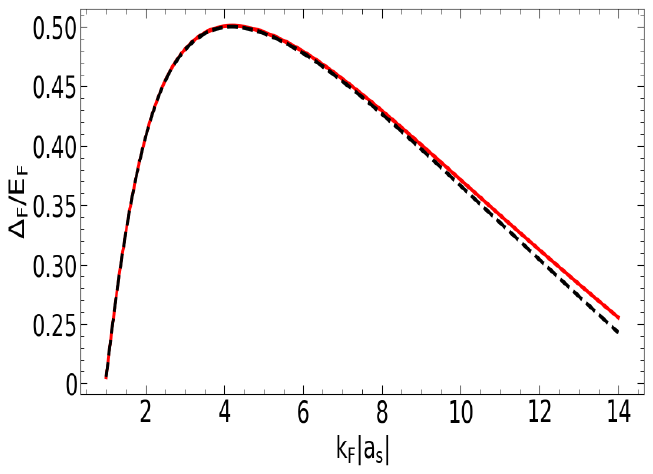}\hspace{0.01in}
\vspace*{-11.9cm}
 \caption{Solid red: BCS predictions by the N$^3$LO(450) 2NF. Dashed black: 2NF plus 3NF up to N$^3$LO.}
\label{fig6}
\end{figure*}

In Fig.~\ref{fig5}, the singlet gap, calculated as described above, is shown in units of the Fermi energy as a function of $k_F|a_s|$. For the purpose of covering a broad spectrum of (bare) interactions, predictions are given for the chiral potential, the meson-exchange CD-Bonn potential, and a local phenomenological potential of the past~\cite{Reid}. While the first two cases are very close, the older local potential generates noticibly lower values of the gap, due to weaker attraction, typical of local interactions, in $^1S_0$. The dots are from QMC calculations of cold atoms~\cite{Gan22}. Neutron matter displays  universal behavior up to values of $k_F|a_s|$ approximately equal to 4, at which point it departs from the unitary gas, indicating that finite-range effects are no longer negligible.

Inclusion of 3NFs and other medium effects results in considerable, sometimes extreme, model dependence~\cite{Gan22, Driss22, Pav17, Benh17}. In Fig.~\ref{fig6}, our basic BCS predictions with N$^3$LO(450) are compared with those where both the leading and subleading 3NF are included. At the low densities considered in the figure, (up to 9\% of saturation density), we find the effect to be a mild reduction.

\section{Conclusions, work in progress, and future plans }                                                                  
\label{Concl}

Dilute NM displays interesting features due to its similarity with a unitary Fermi gas. To explore those features in more depth, we constructed nucleon-nucleon potentials with nearly infinite $^1$S$_0$ neutron-neutron scattering length. We calculated the Berstch parameter and compared with other predictions obtained with ultra-cold atomic gases. We noted that a small tuning of the leading order contact parameter in $^1S_0$ can produce huge variations in the singlet scattering length.

We discussed the unitarity regime as a significant constraint to normal density neutron matter, the symmetry energy, and its density dependence. We took the opportunity to underline the importance of a realistic description of low-energy few-nucleon systems for robust predictions of neutron and neutron-rich matter.

We then moved to a discussion of the singlet pairing gap and presented our baseline results, on which we are building systematic calculations including other medium effects besides 3NFs, such as selfconsistently determined single-particle energies.

In the inner crust of neutron stars, neutron-rich nuclei are arranged in a crystal lattice surrounded by neutrons in a superfluid state.
At densities up to one-half of saturation density, these unbound neutrons form Cooper pairs in the dominant $^1$S$_0$ state, strongly attractive 
at low momenta. 
The physics of the inner crust is largely dependent on this S-wave neutron superfluid, which has been observed through pulsar glitches and modifications of the neutron star cooling due to strong modifications in neutrino emission.

Generally, calculations beyond BCS show a considerable spread of predicted values, depending on which and how medium modifications are included. In closing, we note that, for the gap in the triplet channel, the status is much more unsettled. For example, in Ref.~\cite{Papa17}, using two parametrizations of the UIX 3NF the authors find conflicting answers to whether a triplet P-wave gap even exists.
We are currently working on systematic predictions of the neutron-neutron pairing gap in the $S$- and $P$-wave channels.

\section*{Acknowledgments}
This work was supported by 
the U.S. Department of Energy, Office of Science, Office of Basic Energy Sciences, under Award Number DE-FG02-03ER41270. I am grateful to A. Gezerlis and G. Palkanoglou for helpful discussions on the pairing gap and the BCS equation.

%\reftitle{References}

\end{document}